\documentclass[aps,twocolumn]{revtex4}
\def\BibTeX{{\rm B\kern-.05em{\sc i\kern-.025em b}\kern-.T
    08em\kern-.1667em\lower.7ex\hbox{E}\kern-.125emX}}
\usepackage{graphicx}

\begin{document}

\title{Imprinting skyrmions in thin films by ferromagnetic and superconducting templates}
 

\author{Nuria Del-Valle, Sebastia Agramunt-Puig, Carles Navau, and Alvaro Sanchez$^*$}
\affiliation{Departament de F\'{\i}sica, Universitat Aut\`onoma de Barcelona, 08193 Bellaterra,
Barcelona, Catalonia, Spain}

\begin{abstract}

Magnetic skyrmions are promising candidates as information carriers in a new generation of memories. How to generate and stabilize skyrmions is essential for their successful application to technology.  Here we theoretically demonstrate that arrays of skyrmions can be imprinted in ultrathin ferromagnetic films in large numbers by bringing a magnetic nanostructured template close to the film. Two kind of templates, allowed by present-day nanotechnologies, are studied: arrays of ferromagnetic nanorods and superconducting vortices.  Skyrmions are generated when exposing magnetic films to the template fields for short times and remain stable after removing the template.

\end{abstract}

\maketitle

\begin{figure*}[htb]
\centering

\includegraphics[width=1.0\textwidth]{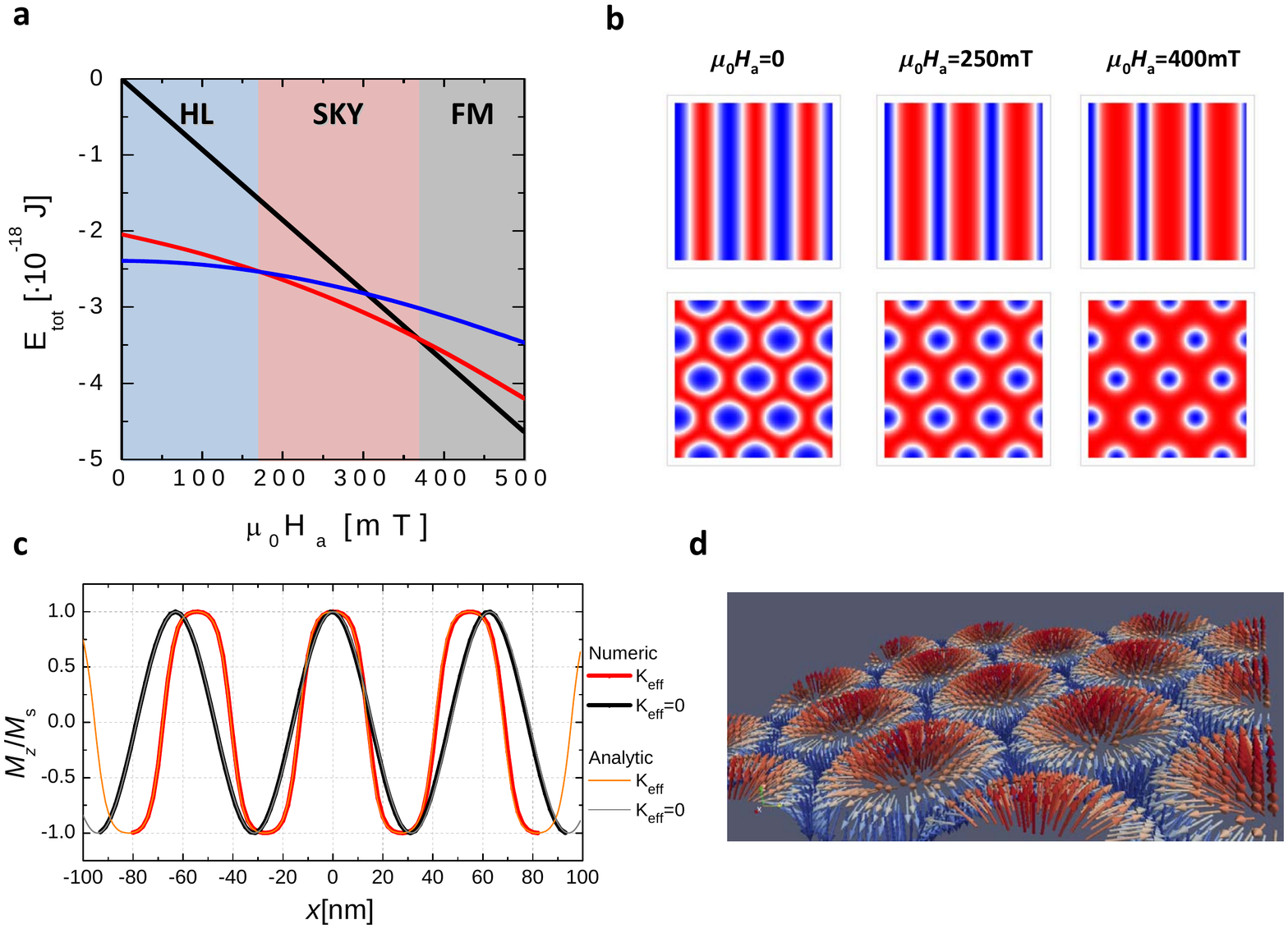}
\caption{
Calculated micromagnetic energy (including the DM, exchange, and Zeeman energies) of a 0.4nm-thick and 200nm-side sample with in-plane periodic conditions for a helical (HL, blue line), skyrmion lattice (SKY, red line) and ferromagnetic (FM, black line) states as a function of the out-of-plane applied magnetic field at 0K (see Methods for the used parameters). 
b) Magnetization distribution of the HL (upper row) and SKY lattice (lower row) states in (a) for different applied fields. In the color scale, red, blue and white represent $M_z/M_s$=1, -1, and 0, respectively. 
c) Comparison between analytical and numerical calculations of the magnetization profile of an infinite strip at zero applied field and 0 K for the cases with effective perpendicular anisotropy $K_{\rm eff}=0$ (black) and $K_{\rm eff}=0.6$MJ/m$^3$ (red).
d) Schematic representation of the obtained hedgehog skyrmion lattice. The arrows represent the magnetization direction.}
\end{figure*}

Skyrmion lattices of nanometer range have been observed in magnetic thin films induced by interfacial Dzyaloshinskii-Moriya (DM) interactions \cite{Ferriani,Heide}, following the lattices initially observed in bulk non-centrosymmetric crystals \cite{Muhlbauer,Yu}. Some important parameters can be adjusted experimentally in thin films, such as the perpendicular magnetic anisotropy \cite{chappert} or the DM strength \cite{Sampaio}. This makes magnetic films, such as Co layers on Pt, excellent systems for studying and optimizing skyrmions \cite{Sampaio}.

The potential use of magnetic skyrmions as information carriers \cite{fert_comment,felser,nagaosa} would require the controlled nucleation of a large number of them. Recently,  
skyrmions have been experimentally generated by processes such local injection of currents \cite{Sampaio}. In this work we theoretically demonstrate an alternative way to create stable tunable arrays of large numbers of skyrmions in a thin film surface: imprinting them with nanostructured magnetic templates. We show how arrays of magnetic dipoles or monopoles, separated a distance of some tenths of nanometers, can be used as nanomagnetic templates. We discuss below how present nanotechnologies involving either ferromagnetic rods or superconducting vortices can be tailored to fabricate these nanotemplates. 

Non-trivial magnetic structures such as skyrmions appear in ferromagnetic films typically because of the competition between the exchange and the DM interactions \cite{felser}. To simulate these systems, we have developed an own-coded micromagnetic procedure \cite{APL}, extended here to include the additional DM term \cite{Rohart}. The Landau-Lifshitz-Gilbert equation is solved by a 4th order Runge-Kutta method. We consider in-plane periodic boundary conditions to avoid edge effects \cite{Ohe}; all calculations have been checked to represent an extended film, independently of the calculation window. We assume ultra thin samples and typical realistic parameters corresponding to perpendicular magnetized Co layer on Pt \cite{Sampaio} (see Methods for details). Magnetostatic energy is disregarded; we have numerically confirmed that it is negligible in these thin systems (as in \cite{Sampaio}).

\begin{figure*}[ht]
\centering

\includegraphics[width=1.0\textwidth]{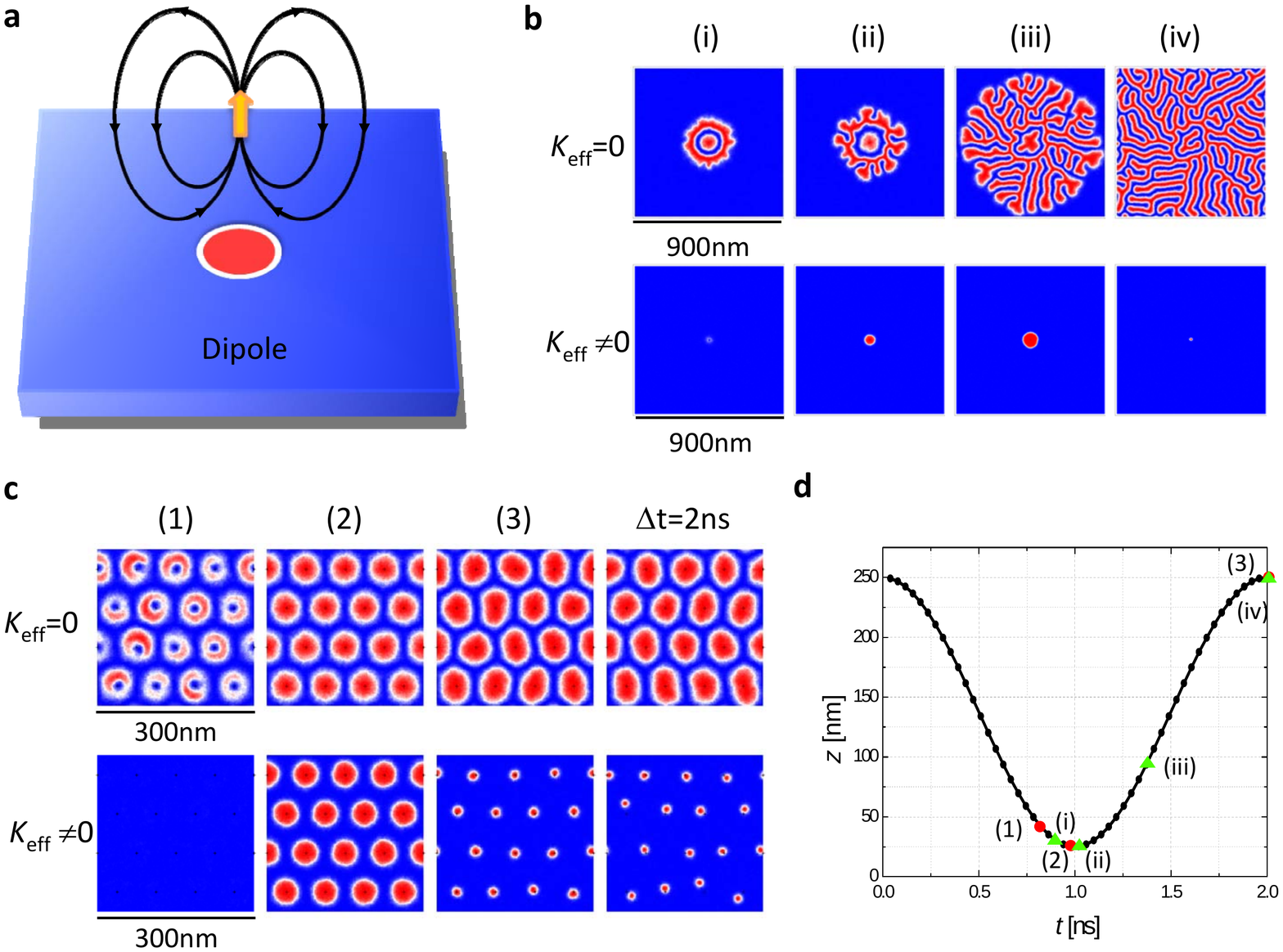}
\caption{
a) Sketch of the magnetization distribution when a magnetic dipole is approaching the thin film in the FM state. 
b) Magnetization distribution of a 0.4nm-thick and 900nm-side sample with in-plane periodic conditions and with $K_{\rm eff}=0$ (upper row) and $K_{\rm eff}=0.6$MJ/m$^3$ (lower row) when a magnetic dipole of magnetic moment $m_d=3\cdot10^{-16}$A/m$^2$ is approaching the sample from a distance of 250nm (dipole magnetic field negligible) to 25nm and then is moved away at 250nm. This process is carried out in 2ns. 
c) Magnetization distribution of a 0.4nm-thick and 300nm-side sample with in-plane periodic conditions and with $K_{\rm eff}=0$ (upper row) and $K_{\rm eff}\neq0$ (lower row) when an array of magnetic dipoles of magnetic moment $m_d=3\cdot10^{-16}$A/m$^2$ is approaching the sample following the same process as in (b). The right column corresponds to the magnetization distribution 2ns after that the array is removed. The distance between the dipoles of the array is 75nm in both $x$ and $y$ directions.
In the color scale of (a), (b) and (c), red, blue and white represent $M_z/M_s$=1, -1, and 0, respectively. Crosses in (b) and (c) indicate the position of the magnetic dipoles.
d) Evolution of the vertical distance between the sample and the single dipole or the dipole array. The points (I), (II), (III) (green triangles) correspond to the plots of (b) and (1), (2), (3) (red dots) correspond to the ones of (c).}

\end{figure*}

\begin{figure*}[ht]
\centering
\includegraphics[width=1.0\textwidth]{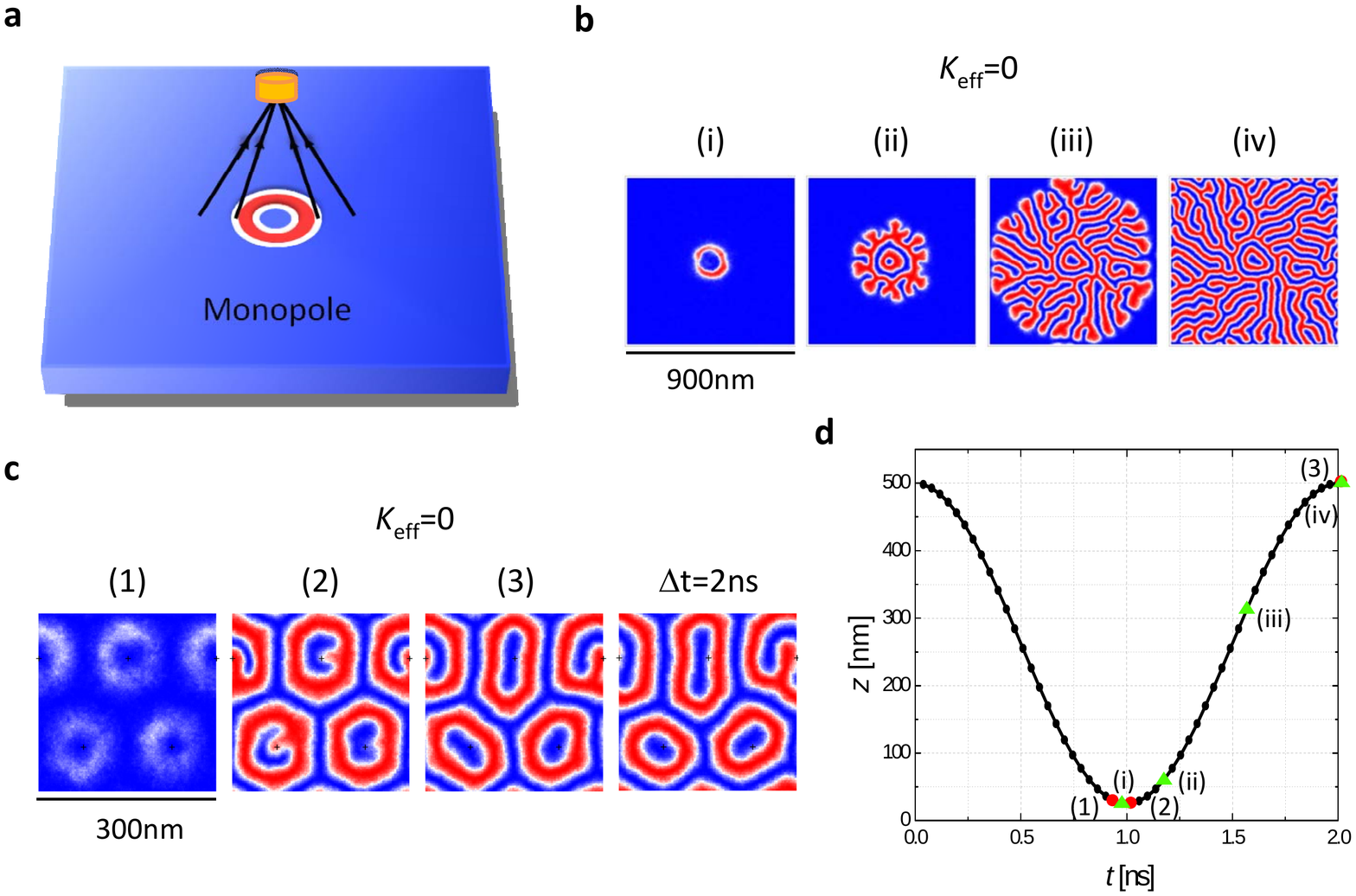}
\caption{
a) Sketch of the magnetization distribution when a magnetic monopole is approaching to the thin sample in the FM state. 
b) Magnetization distribution of a 0.4nm-thick and 900nm-side sample with in-plane periodic conditions and with $K_{\rm eff}=0$ when a magnetic monopole of charge $2\Phi_0=4\cdot 10^{-15}$Wb is approaching the sample from a distance of 500nm (monopole magnetic field negligible) to 25nm and then is moved away. This process is carried out in 2ns. 
c) Magnetization distribution of a 0.4nm-thick and 300nm-side sample with in-plane periodic conditions and with $K_{\rm eff}=0$ when an array of magnetic monopoles of charge $2\Phi_0=4\cdot 10^{-15}$Wb is approaching to the sample following the same process as (b). The right column correspond to the magnetization distribution 2ns after that the array is removed. The distance between the monopoles of the array is 150nm in both $x$ and $y$ directions.
In the color scale of (a), (b) and (c), red, blue and white represent $M_z/M_s$=1, -1, and 0, respectively. Crosses in (c) indicate the position of the magnetic monopoles.
d) Evolution of the vertical distance between the sample and the single monopole or the monopole array. The points (I), (II), (III) (green triangles) correspond to the plots of (b) and (1), (2), (3) (red dots) correspond to the ones of (c).}

\end{figure*}

\begin{figure*}[ht]
\centering
\includegraphics[width=0.5\textwidth]{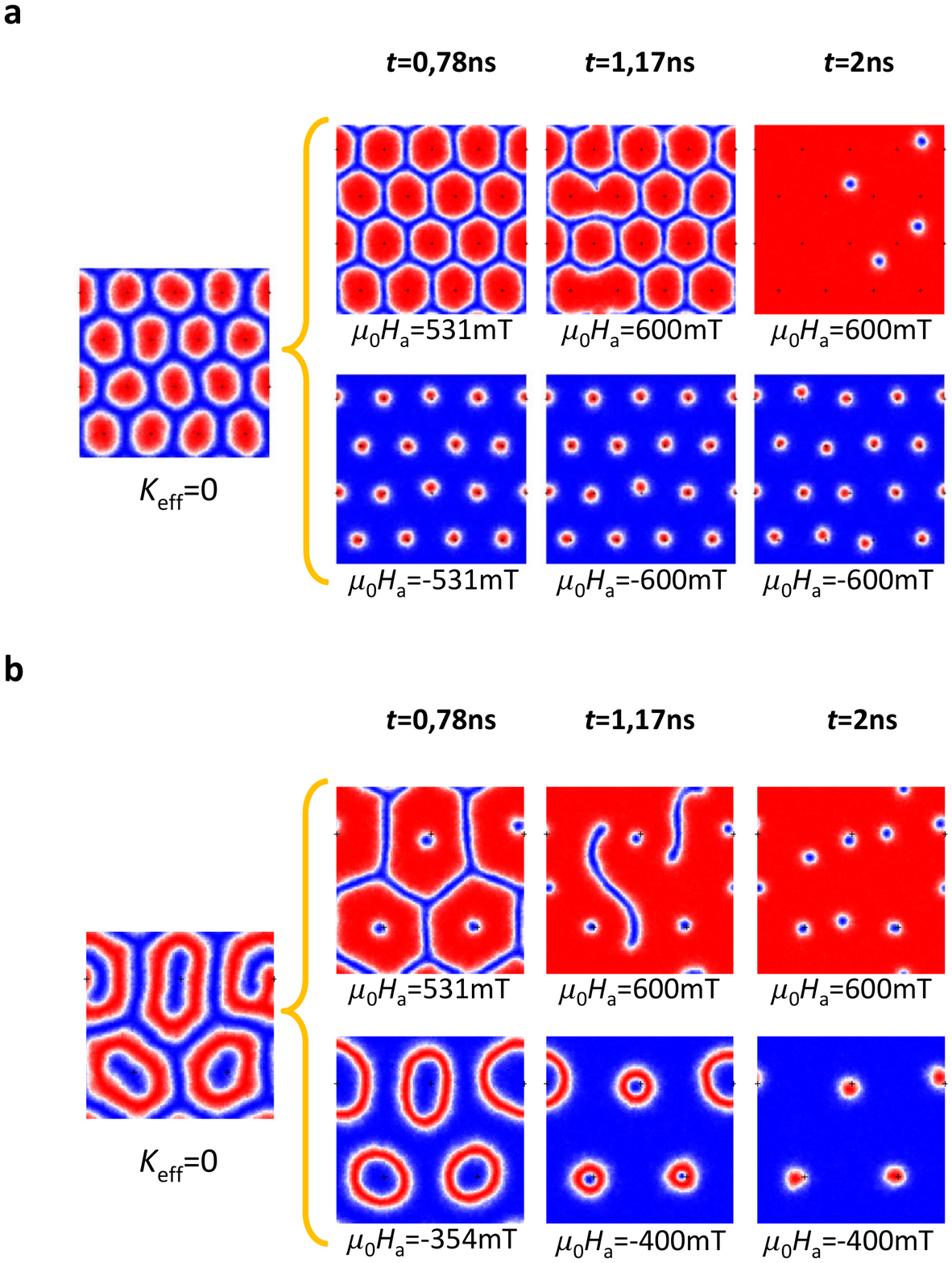}
\caption{
Evolution of some of the states induced by dipoles and monopoles when an external magnetic field is applied. Field is applied during 2ns in two phases: increasing from zero to a maximum value in the first nanosecond and keeping  this maximum field in the second nanosecond.
a) Evolution of the magnetization distribution of the final state (without dipoles) of fig. 2c when a maximum field of 600mT (upper row) and -600mT (lower row) is applied.
b) Evolution of the magnetization distribution of the final state (without monopoles) of fig. 3c when a maximum field of 600mT (upper row) and -400mT (lower row) is applied.
In the color scale of (a) and (b), red, blue and white represent $M_z/M_s$=1, -1, and 0, respectively, and crosses indicate the former positions for the magnetic dipoles and monopoles.
}

\end{figure*}

We first calculate a magnetic phase diagram for the studied films, assuming no anisotropy for simplicity (fig. 1a). The  total energy, consisting of the DM, exchange, and Zeeman terms, is plotted for different initial magnetic configurations as a function of the applied magnetic field $H_{\rm a}$. 
Similar to other studies \cite{Muhlbauer,Yu,nagaosa}, we find that at zero applied field the minimum energy corresponds to a phase of stripes of undulating out-of-plane component of the magnetization, commonly referred to as the helical phase \cite{nagaosa}. We have analytically derived that the minimum of energy for an infinitely long (in the $y$ direction) film with no applied field follows a harmonic magnetization profile
with period $4\pi A/ D$, around 63nm in our studied case (see Supplementary Material Section II, also for the case of a non-zero anisotropy constant $K_{\rm eff}$ or a given $H_{\rm a}$). These solutions are compared in fig. 1c with our numerical simulations for a longitudinal section of our 2D film in the helical phase. Numerical and analytical results perfectly match, confirming our numerical techniques.

An hexagonal lattice of skyrmions appear as a metastable solution at low applied fields $H_{\rm a}$; this phase becomes the ground state at $H_{\rm a}\sim 170$mT. The obtained skyrmions are hedgehog type (fig. 1d) and their size decreases with increasing $H_{\rm a}$, as in \cite{Sampaio}. When $H_{\rm a}$ increases to about $\sim$ 370mT, the ferromagnetic state becomes the lowest energy state.


Our calculations show that it is not easy to generate skyrmions starting from configurations such a ferromagnetic state (as previously seen \cite{milde,nagaosa}), even at the conditions such that a skyrmion lattice has the minimum energy (e. g. applied field of few hundred mT in our case). The reason for that is that generating skyrmions requires overcoming a topological barrier \cite{nagaosa}. It therefore becomes essential to design a procedure to generate skyrmions in these magnetic films under general conditions, even at zero applied field. We propose here a robust method for imprinting skyrmions by bringing a magnetic template -with a field profile that can locally induce the required topological change- close to the film. The initial purely ferromagnetic state will be converted into a skyrmion phase by using magnetic templates with the adequate protocol and conditions.


We have seen above that the combination of DM and exchange energies results in a relevant length scale $\lambda=4\pi A/ D$ at which non-trivial structures would tend to be formed, so we would like our magnetic template fields to vary at this scale. We start with dipolar sources (fig. 2a). It has been shown \cite{ross,pacheco,nikulina,deteresa} that magnetic nanowires can produce a dipolar stray magnetic field in their vicinity, particularly if they have a small aspect ratio. These nanowires can form arrays with separation of tenths of nanometers, at the scale of interest. Magnetic tips were previously proposed to generate skyrmions and bubbles in RKKY systems \cite{kirakosyana}. Also, magnetic vortices have been proposed for creating skyrmions \cite{sun}, although in that case the interfacial exchange energy, and not stray field was the mechanism for skyrmion generation. 

We show in Fig. 2b the effect of approaching a single magnetic dipole of magnetic moment $m=3\cdot 10^{-16} A/m^2$ \cite{Li} to the film up to a distance of 25nm. If there is no anisotropy ($K_{\rm eff}$=0), the dipole induces a ring of opposite magnetization (red), overcoming the topological barrier. The width of this ring is on the order of $\lambda$. When the dipole is removed, a structure consisting of filaments of typical size $\lambda$ develops and eventually propagates throughout all the film (similar structures were experimentally seen in \cite{milde}). Interestingly, if the film has $K_{\rm eff}\neq 0$ (fig. 2b) then the picture changes. A kind of magnetic bubble is formed at the dipole spot but now anisotropy prevents propagation of the filaments. Instead, the bubble shrinks into a stable magnetic skyrmion of smaller size. 
Interesting features appear when considering arrays of magnetic dipoles as templates. In fig. 2c we show that stable arrays of large bubbles (when $K_{\rm eff}=0$) or smaller skyrmions (when $K_{\rm eff} \neq 0$) are created and remain even after removing the template.
These arrays are stabilized owing to the interaction between neighbouring bubbles, via DM and exchange interactions, since magnetostatic energy is negligible. Varying the lattice constant of the dipole arrays results in a rich behavior (Supplementary Material section III). An important consequence of these simulations is that stable skyrmions appear in ultrathin magnetic films with magnetic anisotropy without the need of applying an external overall magnetic field. This happens, both in the the case of a single skyrmion induced by a single magnetic dipole (fog. 2b) and a disordered array of them induced after approaching an ordered array of magnetic dipoles (fig. 2c).


Another interesting possibility to explore is using monopolar field sources (fig. 3a). These can be created either by using arrays of long nanowires \cite{ross,pacheco,nikulina,deteresa} or, alternatively, by arrays of superconducting vortices in films \cite{carneiro,SVL}. The latter can be tailored with separations of few tenths of nanometers by having an array of antidots in superconducting films (\cite{mosh} and references therein). At distance of the order of the Pearl penetration depth (typically, tenths on nanometers \cite{pearl,SVL}), the vortices in thin superconductors create to a good approximation a monopolar field with magnetic charge $Q_0=2\Phi_0$ \cite{carneiro,SVL}, where $\Phi_0$ is the flux quantum. We show in fig. 3b the  magnetization distribution of a film with in-plane periodic boundary conditions when a monopolar field is approached to a distance of 25nm and then is moved away. Similar structures as those for single dipoles (fig. 2b) are found. An array of monopoles create magnetic structures following the monopoles array periodicity, as for dipoles, but the structures have a ring shape of alternating magnetization instead of a bubble structure.
Interestingly, we observe that no magnetic structures can be generated by monopolar fields when $K_{\rm eff}\neq 0$, both for single sources and arrays (although they can be imprinted if the monopoles are closer to the film). The reason for that is that the stray field created by vortex monopoles at 25nm is less than that required to overcome the anisotropy effective field (see Supplementary Material section I), on the order of 2T (whereas dipolar sources at 25nm do provide such field values).



Thus, arrays of dipolar sources generate bubble-like magnetic structures that transform, after some elapsed time, into an array of skyrmions if the film has some perpendicular anisotropy (fig. 2c), even in the absence of an overall applied magnetic field. For monopolar sources, magnetic structures also appear for films with low anisotropy, whereas the topological barrier cannot be overcome in some cases when there is anisotropy. We now demonstrate that, for all cases including films without anisotropy, the arrays of imprinted magnetic structures can be turned into arrays of skyrmions simply by applying a uniform perpendicular magnetic field $H_{\rm a}$. On fig. 4a we show how applying $H_{\rm a}$ results in the presence of skyrmions or anti-skyrmions. When the applied field is positive only a few anti-skyrmions are found in a background of positive magnetization, whereas an array of skyrmions strongly correlated with that of the used imprinting template exist when the applied field is negative. 



Both arrays of ferromagnetic nanowires and superconducting vortices are good candidates for magnetic imprinting also in terms of the stability of their magnetization. We have calculated that the stray field corresponding to the generated magnetic bubbles and skyrmions (see Supplementary Material section IV) at the shortest distance of 25nm is about 1mT, below the coercivity of typical arrays of magnetic nanowires (around 15mT\cite{nikulina,pacheco}). Also, this field is less than that typically needed to depin superconducting vortices in arrays of antidots (around 20mT for a typical Pb film \cite{silhanek}). Thus, the proposed magnetic templates are not demagnetized by the skyrmion's stray fields.


In conclusion, we have shown how templates of nanometer-scale magnetic objects can overcome topological barrier and imprint magnetic structures  in magnetic films that can be turned into arrays of skyrmions. Stable skyrmions can be imprinted in this way in large numbers after expositions as short as few nanoseconds. When using templates creating local dipolar fields (e. g. by ferromagnetic nanorods), skyrmions are imprinted without the need of an overall applied field. This hints at the possibility of using our ideas as a fast way to write large quantities of elements of few nanometers size, which could be manipulate and read through the free surface of the film, for a future generation of magnetic memories.


\section*{Acknowledgements}

We thank Spanish projects NANOSELECT (CSD2007-00041) and MAT2012-
35370 for financial support. AS acknowledges financial support from ICREA Academia (Generalitat de Catalunya) .





\section*{Corresponding author}

Alvaro Sanchez (alvar.sanchez@uab.cat)


\end{document}